\documentclass[journal,article,accept,moreauthors,12pt,a4paper]{mdpi}
\lastpage{x}
\doinum{10.3390/------}
\pubvolume{xx}
\pubyear{2015}
\history{Received: xx / Accepted: xx / Published: xx}

\Title{Short-lived lattice quasiparticles for strongly interacting fluids}

\Author{M. Mendoza $^{1,}$*, and S. Succi $^{2,\dagger}$}

\address{%
$^{1}$ ETH
  Z\"urich, Computational Physics for Engineering Materials, Institute
  for Building Materials, Wolfgang-Pauli-Strasse 27, HIT, CH-8093 Z\"urich
  (Switzerland)\\
$^{2}$ Istituto per
  le Applicazioni del Calcolo C.N.R., Via dei Taurini, 19 00185, Rome
  (Italy),\\and Institute for Advanced Computational Science, Harvard University,
  Oxford Street, 29, MA 0213879, Cambridge, (USA)}

\contributed{$^\dagger$ These authors contributed equally to this work.}

\corres{mmendoza@ethz.ch, +41 44 633 20 93 .}

\abstract{It is shown that lattice kinetic theory based on short-lived quasiparticles proves very effective in simulating the complex dynamics of strongly interacting fluids (SIF). In particular, it is pointed out that the shear viscosity of lattice fluids is the sum of two contributions, one due to the usual interactions between particles (collision viscosity) and the other due to the interaction with the discrete lattice (propagation viscosity). Since the latter is {\it negative}, the sum may turn out to be orders of magnitude smaller than each of the two contributions separately, thus providing a mechanism to access SIF regimes at ordinary values of the collisional viscosity. This concept, as applied to quantum superfluids in one-dimensional optical lattices, is shown to reproduce shear viscosities consistent with the AdS-CFT holographic bound on the viscosity/entropy ratio. This shows that lattice kinetic theory continues to hold for strongly coupled hydrodynamic regimes where continuum kinetic theory may no longer be applicable.}

\keyword{Strongly coupled fluids; Lattice Boltzmann; Quantum Fluids}

\begin{document}


\section{Introduction}

The study of transport properties of strongly interacting fluids (SIF) has gained central stage in modern condensed matter research, with many fascinating connections with quantum-relativistic hydrodynamics, high-energy physics and string theory \cite{SACHbook}. By definition, SIF are moving states of matter in which the interactions are strong to the point of preventing the microscopic degrees of freedom from propagating freely over significant distances as compared to the range of their interactions. More precisely, the mean-free path becomes smaller than the interaction range, and, eventually, the De Broglie length. Remarkable examples in point are found mostly in the framework of quantum fluids, namely quark-gluon plasmas \cite{shuryak2004does}, electrons in graphene \cite{muller2009graphene} and Bose-Einstein condensates in optical lattices \cite{PhysRevA.72.043601}. Note that these extreme and exotic states of matter cover a breathtaking range of densities and temperatures, from one particle/fm$^3$ and trillions Kelvin degrees in ultra hot-dense quark-gluon plasmas, to about $10^{12}$ particles/cm$^3$ and tens of nanokelvins in ultra-cold Bose-Einstein condensates. Thus the SIF regimes show an impressive degree of universality, whose understanding stands as a great challenge at the crossroad of statistical fluid mechanics, condensed matter and high-energy physics. Under SIF conditions, most of the familiar and powerful notions of kinetic theory call for a profound revision. In particular, it is argued that not only Boltzmann's kinetic theory, but the very notion of quasi-particles as weakly-interacting collective degrees of freedom, would loose meaning due their inconspicuously short lifetime \cite{SACHMUL}. Hence, new ideas and methods, both analytical and numerical are in great demand.

Among the analytical ones, a most prominent role has been taken by the AdS-CFT (Anti de Sitter - Conformal Field Theory) duality between gravity in $(d+1)$-dimensions and conformal field theory in $d$-dimensions \cite{MALDA}. The beauty, and practical import, of the dual-holographic picture is that one can solve an otherwise intractable strongly coupled CFT by dealing with its weakly coupled gravitational analogue, an approach sometimes known as {\it holographic principle}. One of the major outcomes of the holographic approach is the so called minimum-viscosity bound \cite{POLISON}
\begin{equation}
\label{MVB}
\frac{\eta}{s} \ge \frac{1}{4 \pi} \frac{\hbar}{k_B} \quad ,
\end{equation}
where $\eta$ is the shear viscosity and $s$ the entropy per unit volume. It has been found that many quantum relativistic fluids, such as quark-gluon plasmas and electrons in graphene, come much closer to matching the above bound than any previously known fluid, including superfluid Helium-III. Moreover, the holographic AdS-CFT picture shows that hydrodynamics continues to apply even though kinetic theory does (may) not. Besides the conceptual challenge of developing statistical mechanics without quasiparticles, the ultra-low values of $\eta/s$ open up the exciting possibility of observing dynamical instabilities and ensuing (pre) quantum turbulence phenomena in these extreme states of matter \cite{MILLER}.

It is hereby maintained that a proper lattice formulation of effective kinetic theory can effectively describe the dynamics of SIF's by making use of very short-lived quasiparticles. The two main conceptual ingredients are as follows: 1) Top-down approach: {\it design} kinetic equation solely based on symmetry and conservation properties of the macroscopic field-theory (hydrodynamics), including entropic constraints. This just the reverse of the canonical route of {\it deriving} kinetic theory from first-principles, i.e. by coarse-graining underlying microscopic dynamics (which is the process generating quasi-particles). 2) The existence of {\it negative} propagation viscosity, as an emergent property due to the interaction of the free-streaming quasiparticles with the discrete lattice. Item 2) is crucial to sustain ephemeral quasiparticles, with lifetimes as small as one millionth of the free-flight time, hence to achieve ultra-low viscosities, well below the natural kinematic lattice viscosity $\nu_L = \Delta x^2/\Delta t$, $\Delta x$ and $\Delta t$ being the lattice spacing and hopping time, respectively.


\section{Results and Discussion}

\begin{figure}
\begin{center}
\includegraphics[width=0.8\columnwidth]{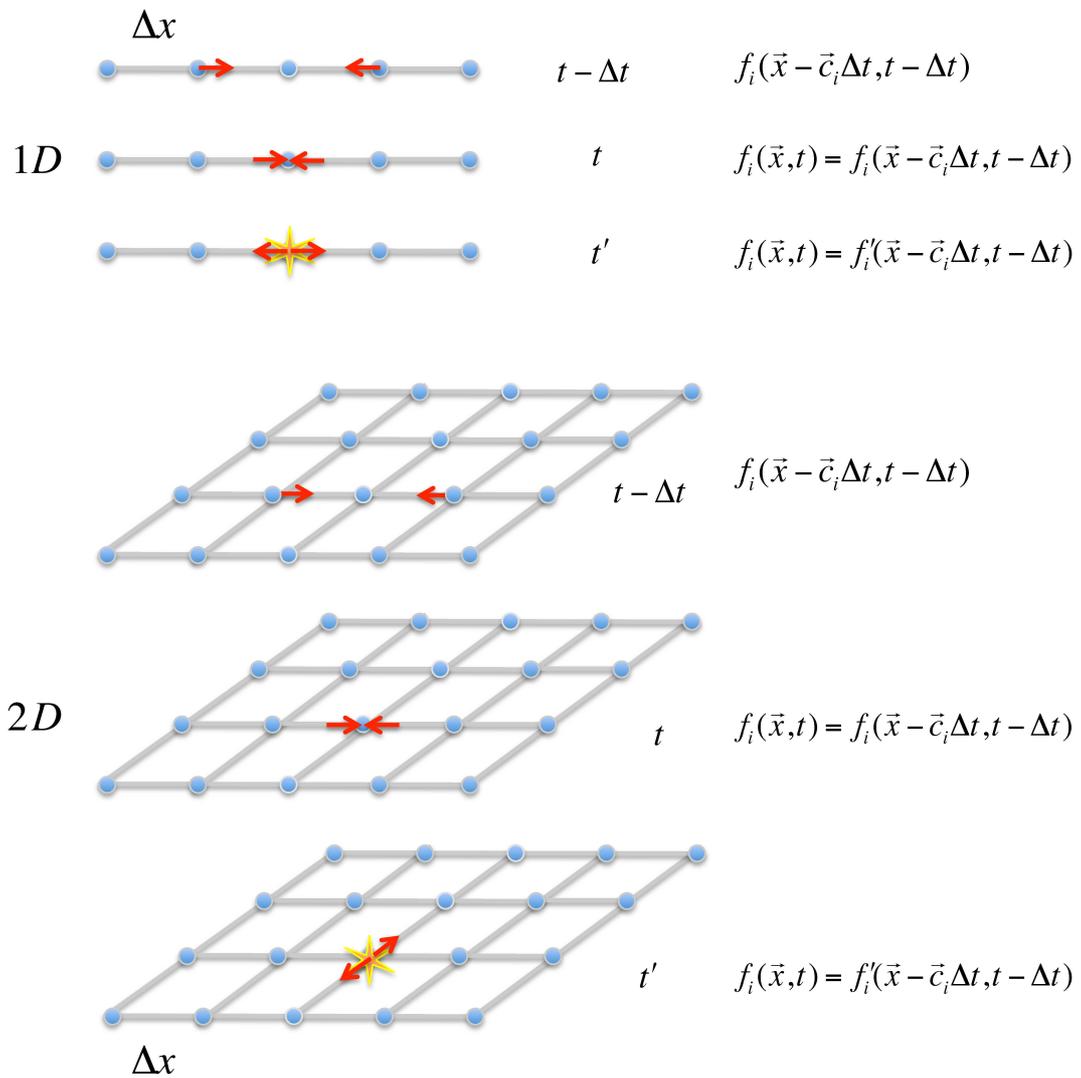}
\caption{Schematics of the LB scheme: the quasi-particle distributions hop along the links defined by the discrete velocities and, once on the same site, they scatter into the post-collisional state. One- (top) and two- (bottom) dimensional lattices. Here $\Delta x$ is the size of the lattice.}
\label{fig1}
\end{center}
\end{figure}

The typical lattice kinetic equation reads as follows \cite{LBE}:
\begin{eqnarray}\label{LB}
f_i(\vec{x}, t)&-f_i(\vec{x}-\vec{c}_i \Delta t,t-\Delta t) = 
-\frac{\Delta t}{\tau} [f_i-f_i^{\rm eq}](\vec{x}-\vec{c}_i \Delta t, t-\Delta t),
\end{eqnarray}
where $f_i(\vec{x}, t)$ is the probability of finding a particle at position $\vec{x}$ in the lattice at time $t$ with discrete velocity $\vec{v}=\vec{c}_i$. The left hand side represents the free-streaming from site to site while the rhs encodes collisional relaxation to a local equilibrium distribution $f_i^{eq}$ on a time scale $\tau$ (see Fig.~\ref{fig1}). The discrete Boltzmann distributions (``populations'') are quintessential quasiparticles, as they encode the dynamics of a collection of molecules in a lattice of size $\Delta x^3$ (in three dimensions), moving along the direction $\vec{c}_i$. The local equilibrium is a universal function  (e.g. Maxwell-Boltzmann, Bose-Einstein, Fermi-Dirac, etc) of the locally conserved quantities (hydrodynamic modes), such as flow density $\rho = m \sum_i f_i$ and current $J=\rho \vec{u}= m \sum_i \vec{c}_i f_i$, and dictates the structure of the inviscid macroscopic equations (Euler). Here $m$ denotes a characteristic mass. The relaxation time, on the other hand, corresponds to the rate at which this equilibrium is reached, and dictates the fluid transport coefficients.

In the Boltzmann kinetic theory, in the strong-coupling regime $\tau \to 0$, viscosity is directly proportional to this relaxation time, $\nu \sim  c_s^2 \tau$, $c_s = \sqrt{k_BT/m}$ being the thermal speed \cite{cercignani1975theory}. In the lattice, however, this expression receives a crucial contribution from the discrete free-streaming. Indeed, by performing the appropriate Chapman-Enskog asymptotic expansion in the Knudsen number $Kn=c_s \tau/L$, $L$ being a typical hydro-scale, one obtains:
\begin{equation}\label{viscosity_eq}
\nu = c_s^2 (\tau-\Delta t /2) \quad ,
\end{equation}
where the negative term at the rhs is the contribution of the free-streaming. Given the minus sign, it is immediately seen that ultra-low viscosities of order $\epsilon \nu_L$ can be achieved by near-matching the free-flight and collisional time, i.e. by choosing  $\tau=\Delta t(1+ \epsilon)/2$. In lattice units, one can write Eq.~\eqref{viscosity_eq} as
\begin{equation}
\label{VISCO}
\nu = {\cal C}_s^2 \nu_L (1/\omega-1/2) \quad ,
\end{equation}
where $\nu_L=\frac{\Delta x^2}{\Delta t}$ is the natural lattice viscosity and we have set $\omega = \Delta t/\tau$. In the above ${\cal C}_s$ is the sound speed in lattice units, typically $1/\sqrt 3$ in most common lattices. Crucial for the correct recovery of the hydrodynamic equations is that the lattice exhibits enough symmetry to ensure the proper conservation laws as well as rotational invariance \cite{FHP}.

To further appreciate the point of ultra-low viscosity, let us now recast the kinetic equation (\ref{LB}) in the most compact stream-collide form: 
\begin{equation}
\label{LBC}
f_i(\vec{x},t) = f'_i (\vec{x}-\vec{c}_i \Delta t, t-\Delta t)\quad ,
\end{equation}
where 
\begin{equation}
\label{LB2}
f'_i  \equiv (1-\omega) f_i + \omega f_i^{eq} \quad ,
\end{equation}
is the post-collisional distribution (see Fig.~\ref{fig1}). 
The latter expression highlights three distinguished updates: i) $\omega=0$, $f'_i=f_i$; The particles are fully uncoupled (no collision) perform ballistic motion and never equilibrate, corresponding to infinite diffusivity. 
ii) $\omega=1$, $f'_i=f_i^{eq}$; The particles reach equilibrium in a single time-step. According to Eq.~\eqref{VISCO}, this gives a viscosity $\nu = \nu_L {\cal C}_s^2/2$, i.e. comparable with the lattice viscosity $\nu_L$. iii) $\omega=2$, $f'_i = f_i - 2 f_i^{neq}$; Collisions send the populations exactly ``on the other side'', to the mirror state $f_i^* \equiv f_i^{eq}-f_i^{neq}$, defined by complete reversal of the non-equilibrium component $f_i^{neq} \equiv f_i-f_i^{eq}$. Based on Eq.~\eqref{VISCO}, this yields formally zero viscosity, i.e. infinitely strong coupling. 

Hence, by choosing $\tau = \Delta t/2$, the quasi-particles have literally zero life-time, even though both free-streaming and collisional time-scales are finite. Far from being a mere mathematical nicety, this regime is crucial to the operation of the LB scheme in the very-low viscous regime relevant, say, to fluid turbulence. Indeed, given the current computer resolutions, iii) is precisely the regime that needs to be approached in order to simulate strongly coupled classical fluids, such as turbulent flows at Reynolds number above a few thousands. To clarify the point, let us recall that the strength of fluid turbulence is measured by the Reynolds number, $Re=UL/\nu$, where $U$ and $L$ are macroscopic velocity and length scales, respectively. In lattice Boltzmann units, $\Delta x=\Delta t = 1$, this gives $Re=Ma N/\nu_{lb}$, where $Ma=U/c_s$ is the thermal Mach Number, $N=L/\Delta x$ is the number of lattice sites per linear dimension and $\nu_{lb}  \equiv \nu/\nu_L=(1/\omega-1/2)$, is the viscosity in lattice Boltzmann units. Current computers allow at best $N \sim 10^4$, so that with Mach numbers of order $1$, reaching $Re=10^7$ (air flowing around a standard car), requires $\nu_{lb} \sim 10^{-3}$, i.e. three orders of magnitude below the natural lattice viscosity $\nu_L$. This simple example highlights the necessity of operating with very short lived quasiparticles, where short means very small as compared to the physical free-streaming and collisional time scales, $\Delta t$ and $\tau$, both $O(1)$ in lattice Boltzmann units.

These quasi-particles function pretty well in the discrete world they live in, if only very shortly. In fact, by promoting the relaxation parameter $\omega$ to the status of a dynamic field responding self-consistently to the local constraints of the second principle (Boltzmann's H-theorem), the lifetime of these ``ephemeral'' quasi-particles can be made as small as one millionth of their natural value, i.e $\omega \simeq 2(1-10^{-6})$ \cite{ELBRMP,ELB1,ELBKC}. In order to appreciate why this is remarkable, let us consider the lattice kinetic update in compact form, Eq.~\eqref{LBC}. The realisability constraint imposes that the discrete populations be non-negative, hence, starting with a non-negative pair $(f_i,f_i^{eq})$, the collisional update should return a non-negative $f'_i$. A moment thought shows that this is indeed the case in the safe under-relaxation regime $0<\omega<1$. However, we have just shown that such regime does not provide access to the large Reynolds numbers typical of most turbulent flows. Hence, the LB update needs to operate {\it deeply} into the potentially unprotected over-relaxation regime, $1<\omega<2$. The remarkable fact is that this is indeed possible, thanks to the existence of a lattice analogue of the H-theorem \cite{ELBRMP,ELB1,ELBKC}. Note that the regime $1<\omega<2$ is called over-relaxation because the time step is larger than the relaxation time, $\tau < \Delta t$, and consequently, the post-collisional distribution is no longer bounded between the pre-collisional one and the local equilibrium, which is a potential threaten to positive-definiteness.

{\subsection{Propagation viscosity in Luttinger liquids}

Let us apply this concept to Luttinger liquids. 
By dealing with a unitary Fermi gas, one should distinguish two different regimes, weak and strong coupling. Since the present work is focused on strongly interacting fluids, we consider the regime of large values of the Lieb-Liniger parameter, $\gamma \equiv m g/(\hbar^2 n) \gg 1$, where $m$ is the atomic mass, $n$ the atomic density, and $g$ the coupling constant \cite{transition_cold}. It proves expedient to define the dimensionless quantity $K \equiv \hbar n\pi/(m c_s)$. 
Note that $\gamma$ and $K$ relate to each other through an expression  
that, for $1 < \gamma < 10$,  takes the form $K \simeq \pi \left[\gamma - (1/2\pi) \gamma^{3/2}\right]^{-1/2}$ \cite{transition_cold}. 

In the strongly interacting regime and for weak external potentials, a quantum phase transition from Mott-insulator to superfluid occurs for $\gamma > \gamma_c$ and $K_c = 2$ \cite{transition_cold, pinning}. 
\begin{figure}
\begin{center}
\includegraphics[width=0.7\columnwidth]{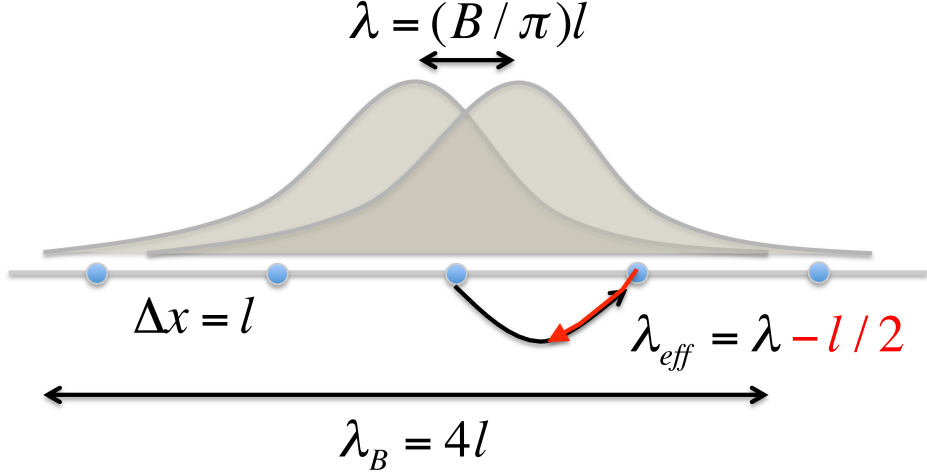}
\caption{Representation of the Luttinger liquid. 
Shown are the mean-free-path $\lambda$, the De Broglie length $\lambda_B$, the lattice size $\Delta x = l$, and the effective mean-free-path predicted by the lattice $\lambda_{eff}$. 
Note that, due to the negative contribution ($-l/2$), lattice kinetic 
fluids can attain ultra-low viscosity even though the standard collisonal
viscosity is of the same order of the lattice viscosity $\nu_L$.}
\label{fig2}
\end{center}
\end{figure}

Let us next define, from Eq.~\eqref{viscosity_eq}, $\nu_c = c_s^2 \tau$ and $\nu_p = c_s^2 \Delta t/2$, denoting ``collision'' and ``propagation'' viscosities, respectively. In order to estimate the propagation viscosity for a SIF at quantum criticality, we recast Eq.~\eqref{viscosity_eq} in the form:
\begin{equation}
\nu_p/\nu_c = \Delta t/2\tau \quad .
\end{equation}
Assuming that $\Delta t = l/c$, where $l$ is the lattice periodicity, with $c$ the light speed, and $\tau = \lambda/c_s$, with $\lambda$ the characteristic mean-free path, the above relation reads:
\begin{equation}
\nu_p/\nu_c =  (1/2) (c_s/c) (l/\lambda) \quad .
\end{equation}

In the sequel, we shall take $c \simeq c_s$. Since at criticality, $K_c = 2$, taking into account that the density $n \simeq 1/l$ (commensurate density \cite{transition_cold}), it follows that $\lambda_B = 4 l$, where $\lambda_B$ is the De Broglie length. Inserting this into the viscosity ratio, we obtain
\begin{equation}\label{visco_crit}
 \frac{\nu_p}{\nu_c} \bigg |_{crit} =  \frac{1}{8} \frac{\lambda_B}{\lambda}.
\end{equation}
Note that SIFs regimes are characterized by the condition $\lambda < \lambda_B$. To determine the ratio $\lambda_B / \lambda$ of the lattice fluid, we follow the same procedure adopted in Ref.~\cite{ultra_gas} to obtain the AdS-CFT viscosity bound. Based on the Heisenberg principle,  $k_B T \tau > \hbar/2$, so that $\tau = \lambda/c = \hbar B /(2k_B T)$, $B > 1$ being a constant. 

Consequently, 
\begin{equation}\label{ratio_lambdas}
    \lambda = \frac{B}{4\pi} \lambda_B \quad .
\end{equation}
The relations between the lattice size $l$, the mean-free path $\lambda$, the De Broglie length $\lambda_B$ and the predicted lattice mean-free path $\lambda_{eff}$, can be appreciated from Fig.~\ref{fig2}. 

Note that the constant $B$ is related to the AdS-CFT bound by \cite{PhysRevA.72.043601},
\begin{equation}
    \frac{\eta}{s} = \frac{B}{4\pi} \frac{\hbar}{k_B} \quad .
\end{equation}
Inserting Eq.~\eqref{ratio_lambdas} into Eq.~\eqref{visco_crit}, we obtain 
\begin{equation}\label{this_one}
 \frac{\nu_p}{\nu_c} \bigg |_{crit} =  \frac{\pi}{2 B} \quad .
\end{equation}
This result calls for a number of comments. 
First, we note that the condition of non-negative viscosity,  $\nu_p /\nu_c \leq 1$, implies $B \geq \pi/2 \simeq 1.6$, which corresponds to a bound $\eta/s \simeq 0.13 \hbar/k_B$. In this case, the lattice quasi-particles can model even more "perfect" fluids than the Luttinger liquid. On the other hand, if we assume that $\tau = \Delta t$, which means that the quasi-particles thermalise in a single time step, we obtain  $B = \pi$, yielding a bound $\eta/s \simeq 0.25 \;\hbar/k_B$, which is very close to the value measured in experiments for two and three dimensional Fermi gases at unitarity \cite{strongfluid}. Note that in order to get Eq.~\eqref{this_one} we have not only used lattice kinetic theory but also the Heisenberg principle, which shows that we still need quantum mechanics contributions for this particular example. However, this does not imply that lattice kinetic theory cannot handle quantum systems by itself. In fact, a quantum lattice Boltzmann model exists since long \cite{QLB} to solve the Dirac equation. It is based on the same stream-collide paradigm of the classical lattice Boltzmann equation, although with a skew-symmetric collision matrix, where $\tau$ is related to $\hbar$ via $\tau= \hbar/mc^2$. 



\section{Conclusions}

In summary, one can conclude that lattice kinetic schemes endowed with an H-theorem support ultrashort-lived computational quasiparticles, down to viscosities several orders of magnitude below their natural scale (the lattice viscosity). This is a strict consequence of the fact that, due the discrete nature of the lattice, free-streaming also contributes a viscosity and, most importantly, a negative one. 
It is as if, in the process of hopping from site to site, the lattice quasiparticles would ``surrender'' part of their collisional viscosity to the lattice itself, and precisely in proportion of half of the flight-time. This is plausible, since negative viscosity corresponds to dynamic instability, namely the system moving away from local equilibrium, a sort of backward move in time (rejuvenation), as opposed to the standard forward move due to collisions. It is also plausible to speculate that this ultra-low viscosity could be regarded as an effect of lattice criticality, in the sense that the two competing processes, collisions and streaming, come to a near exact balance through the interaction of the quasiparticles with the lattice. 

This process follows in the footsteps of similar phenomena in condensed matter: electrons in graphene behave like effective near mass-free excitations precisely because, due to the special symmetry of the honeycomb lattice, the electrons near the Dirac point ``release'' most of their mass to the lattice \cite{graphene1}. Since mass dictates the (Compton) collision frequency, $\omega_c=mc^2/\hbar$, and the collisional viscosity is inversely proportional to $\omega_c$, it is clear that negative viscosity and near mass-free electrons in graphene belong somehow to the same family of lattice-induced phenomena. In the continuum, $\Delta t/\tau \to 0$, the negative viscosity is totally negligible, but in the discrete world it becomes crucial to attain ultra-low kinematic viscosities. 

By using lattice kinetic theory, one can indeed achieve zero viscosity, and the same can be done in the continuum theories, such as continuum Boltzmann and Navier-Stokes equations. However, when the physical system possesses intrinsic discrete properties, e.g. quantum liquids in optical lattices, the viscosity can take very small values due to the presence of the negative viscosity induced by the lattice. For these particular systems, lattice kinetic theory may describe better the transport properties than continuum kinetic theory, as is the case of the lower bound in the shear viscosity of quantum liquids, which is overestimated by the continuum Boltzmann equation. On the other hand, in the classical context, a perfect fluid has zero viscosity, which corresponds in our case to $\omega = 2$. Thus, in the continuum limit, the lattice kinetic theory becomes just a numerical tool and recovers the Euler equations. However, in strongly coupled fluids, which are the focus of this paper and usually are governed by quantum mechanics, fluids do not have zero but a finite value of viscosity. Here we show that considering $\omega = 1$ one can get instantaneous relaxation and at the same time a lower bound for the viscosity of the physical system.

A few concluding remarks regarding strongly coupled fluids are in order. One may argue that most of these fluids are quantal, hence not captured by classical lattice kinetic models, for instance, the LB schemes discussed in this work. To this regard, it is worth pointing out that the lattice Boltzmann method has been already used to study two-dimensional Fermi gas at unitarity \cite{brewer2015lattice}, providing a procedure to calculate transport coefficients precisely. Furthermore, LB versions for quantum wave functions are available since long, and they are based exactly on the same stream-collide paradigm, if only with a suitable scattering matrix for collisions \cite{QLB}. Incidentally, such quantum LB versions provide exact realisations of the  Dirac equation in one spatial dimension and can be extended to $d>1$ by proper operator splitting \cite{FILLI}. Finally, we wish to point out that matrix extensions of the quantum LB scheme have been recently connected with the Hubbard model for strongly correlated quantum fluids \cite{LBUB}, thereby lending further weight to the main idea proposed in this paper, namely that {\it lattice} kinetic theory may provide a new angle of investigation for strongly-interacting fluid regimes where {\it continuum} kinetic theory may no longer hold. 


\acknowledgments{Acknowledgments}

The authors are grateful to Ilya Karlin for many discussions over the years. SS wishes to acknowledge Juan Maldacena for valuable exchanges on the AdS-CFT duality. We acknowledge financial support from the European Research Council (ERC) Advanced Grant 319968-FlowCCS.  


\authorcontributions{Author Contributions}

All authors conceived and designed the research, analysed the data, worked out the theory, and wrote the manuscript.


\conflictofinterests{Conflicts of Interest}

The authors declare no conflict of interest.

\bibliography{biblio}
\bibliographystyle{mdpi}


%


%

\end{document}